\documentclass[aps,prl,groupedaddress,superscriptaddress,twocolumn,notitlepage,bibnotes]{revtex4-1}
\usepackage{graphicx}
\usepackage{dcolumn}
\usepackage{color}
\usepackage{amsmath,amssymb,bm,array,tabularx,booktabs,multirow}
\usepackage{longtable}
\usepackage{braket}
\usepackage{here}
\usepackage[normalem]{ulem}

\begin{document}

\setlength{\extrarowheight}{4pt}

\title{Shear-Strain Controlled High-Harmonic Generation in Graphene}

\author{Tomohiro Tamaya}
\email{tamaya@g.ecc.u-tokyo.ac.jp}
\affiliation{Institute for Solid State Physics, University of Tokyo, Kashiwa, 277-8581, Japan}
\affiliation{JST, PRESTO, 4-1-8 Honcho, Kawaguchi, Saitama, 332-0012, Japan}

\author{Hidefumi Akiyama}
\affiliation{Institute for Solid State Physics, University of Tokyo, Kashiwa, 277-8581, Japan}

\author{Takeo Kato}
\affiliation{Institute for Solid State Physics, University of Tokyo, Kashiwa, 277-8581, Japan}
\date{\today}

\begin{abstract}
We propose a novel method for controlling the high-harmonic generation (HHG) with a high dynamic range in single-layer graphene. We find that, by utilizing shear strain, a significant enhancement or quenching of HHG is possible over a range of several orders of magnitude. This feature is made possible by the resonance mechanism at a van Hove singularity. Therein, the shear strain controls the configurations of the two Dirac cones, resulting in changes in the energy and dipole moment at the saddle point of the band dispersion. Our findings provide a way for modulating or switching light by using a nano-optomechanical device composed of single-layer graphene.
\end{abstract}

\maketitle

Two-dimensional (2D) Dirac electrons in single-layer graphene have attracted much attention~\cite{novoselov2004electric,RevModPhys.81.109} for their novel linear and nonlinear optical properties~\cite{nair2008fine,ando2002dynamical,PhysRevLett.96.256802,zhu2014optical,PhysRevLett.101.196405,PhysRevLett.103.186802,cai2015tunable,falkovsky2007space,baudisch2018ultrafast,bonaccorso2010graphene,sun2010graphene,bao2009atomic,vermeulen2018graphene,sun2010stable,martinez2011mechanical,ma2012graphene,lui2010ultrafast,kim2018ultrafast,xia2009ultrafast,gan2013chip,liu2011graphene,li2014ultrafast,PhysRevLett.105.097401,soavi2018broadband,Yoshikawa2017,chang2010multilayered,cizmeciyan2013graphene,kim2018ultrafast,lui2010ultrafast,gan2013chip,li2014ultrafast,denk2014exciton}. One-atom-thick graphene exhibits universal frequency-independent light absorption, $\pi\alpha=2.3\,{\rm \%}$, where $\alpha=e^2/\hbar c \simeq1/137$ is the fine structure constant~\cite{nair2008fine,ando2002dynamical,PhysRevLett.96.256802,zhu2014optical,PhysRevLett.101.196405,falkovsky2007space,PhysRevLett.103.186802,cai2015tunable}. Ultrafast carrier dynamics on massless bands provide a variety of high-speed broadband optical responses~\cite{baudisch2018ultrafast,bonaccorso2010graphene} which include saturable absorption enabling stable passive mode-locking~\cite{sun2010graphene,bao2009atomic,sun2010stable,vermeulen2018graphene,ma2012graphene,martinez2011mechanical}, light emission~\cite{kim2018ultrafast,lui2010ultrafast}, detection~\cite{xia2009ultrafast,gan2013chip}, and modulation~\cite{li2014ultrafast,liu2011graphene}. Giant optical nonlinearity~\cite{PhysRevLett.105.097401} and high-harmonic generation (HHG)~\cite{soavi2018broadband,Yoshikawa2017} have also been reported, which indicate intriguing strong light-matter interactions in graphene in the visible, infrared and terahertz (THz) regions.

Further studies have targeted artificial control, modulation, or switching of these properties for potential applications in graphene-based optical devices~\cite{soavi2018broadband,Yoshikawa2017,liu2011graphene,otsuji2012graphene,zhao2016review,singh2011graphene,wen2014graphene,grigorenko2012graphene}. Indeed, modulation of the HHG intensity has been demonstrated over a range of two orders of magnitude via tuning of the Fermi energy by gate voltage~\cite{soavi2018broadband} or via tuning of the incident-light ellipticity~\cite{Yoshikawa2017}. However, the robustness of optical transition probability $\pi\alpha$ inherent to 2D Dirac electrons makes the external control of optical responses highly challenging, and a significant modification of the Dirac band dispersion should be necessary to achieve giant switching of optical responses.

A graphene sheet has high flexibility and a shear strain of up to $27\,{\rm \%}$ can be reversibly applied to it~\cite{peng2020strain,frank2011raman,castellanos2015mechanics,bertolazzi2011stretching,castellanos2012elastic,shi2019strain,PhysRevLett.103.046801,ribeiro2009strained,zhao2013review,cadelano2009nonlinear,lee2008measurement,liu2007ab,ni2008uniaxial,PhysRevB.81.081407,bae2013graphene,tian2014scalable,guinea2012strain,PhysRevB.80.045401,gui2008band,PhysRevB.81.241412,si2016strain,li2010strain,PhysRevB.84.245444}. Notably, a few theoretical studies~\cite{PhysRevB.81.241412,si2016strain,li2010strain,PhysRevB.84.245444} have pointed out that shear strain in graphene makes the two Dirac cones at the $K^{\pm}$ points to get closer and finally merge to form a band gap. This strain effect should be useful for control of the nonlinear optical responses of graphene. 

In this paper, we theoretically demonstrate a significant enhancement or quenching of HHG over several orders of magnitude in graphene by tuning the shear strain and the incident-light polarization angle. We clarify that the enhancement stems from the resonance between the incident-light photon energy and the band gap or saddle-point energies at the intermediate point of the two Dirac cones. High-contrast control of the nonlinearity suggests the possibility of using HHG-light switching and modulation in graphene photonics and optoelectronics.

\begin{figure*}[t]
\begin{center}
\includegraphics[width=17.8cm]{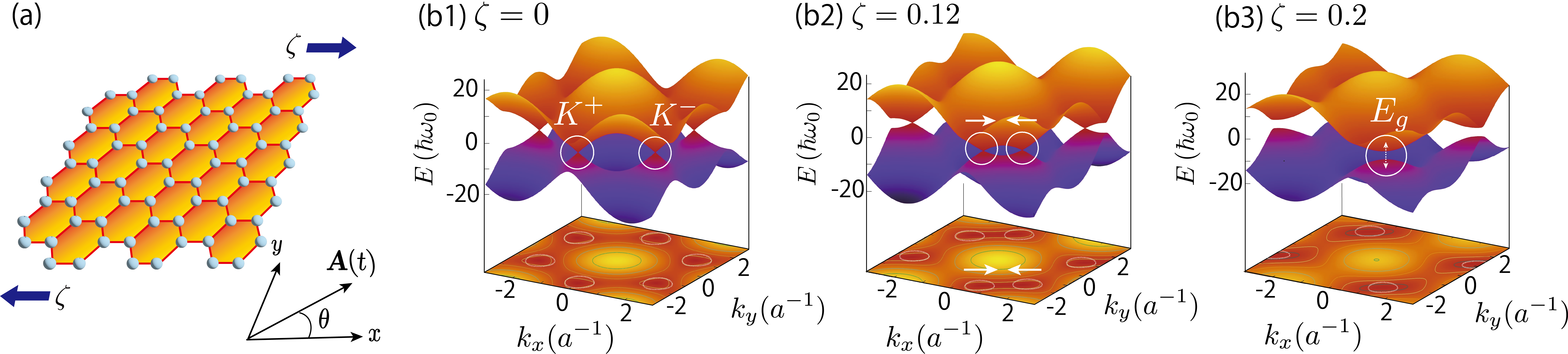}
\caption{(Color online) (a): Schematic figure of lattice structure of graphene under shear strain. The amplitude of the shear strain is represented by $\zeta$ and the vector potential of the incident light is denoted as $\bm{A}(t)$, whose tilt angle measured from the armchair axis is described by $\theta$ (see the inset). (b1)-(b3): Variation in band structures of graphene for the three shear strain parameters, (b1) $\zeta=0$, (b2) $\zeta=0.12$, and (b3) $\zeta=0.2$. The white arrows in Fig.~(b2) indicate the direction of the change in position of the Dirac points induced by the strain.
\label{fig:config1}}
\end{center}
\end{figure*}

\begin{figure*}[bt]
\begin{center}
\includegraphics[width=17.8cm]{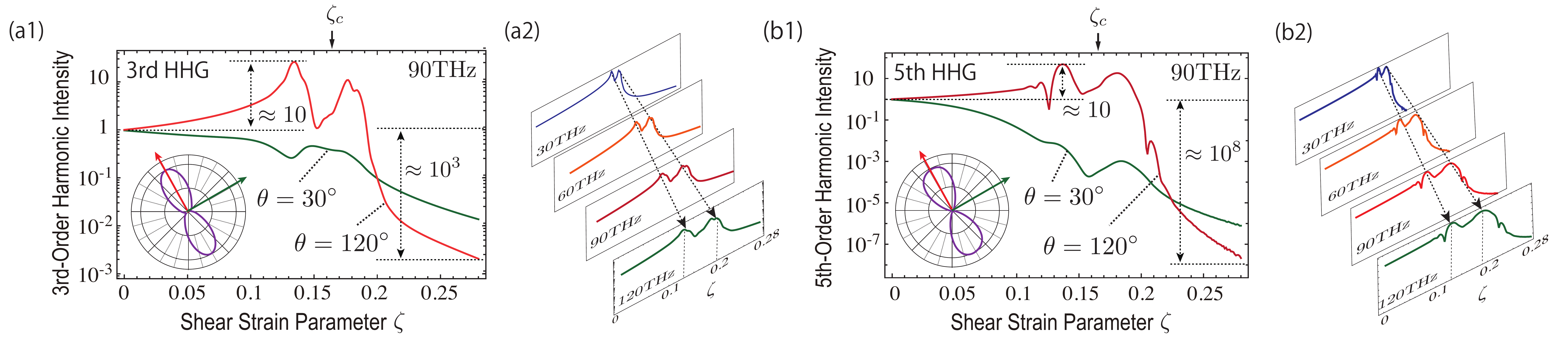}
\caption{(Color online) (a1) and (b1): Shear strain dependence of the third- and fifth-order harmonics for $\omega_{0}=90\rm{THz}$. The red and green lines indicate the high-harmonic intensities for the incident-light angles $\theta=120^{\circ}$ and $\theta=30^{\circ}$, respectively. The polarization angle dependences of the third and fifth harmonics for $\zeta=0.12$ are shown in the insets. {The critical strength of the shear strain for a band gap to form is described by $\zeta_{c}$.} (a2) and (b2): Shear-strain dependence of the third- and fifth-order harmonics for the different incident-light frequencies, 30 THz (blue lines), 60 THz (orange lines), 90 THz (red lines), and 120 THz (green lines). The dotted black arrows indicate the variation in the double peaks.
\label{fig:config2}}
\end{center}
\end{figure*}

The graphene under shear strain can be modeled by a tight-binding model based on a distorted hexagonal lattice (see Fig.~\ref{fig:config1}~(a)). The anisotropic hopping energies are described as $\gamma_{i} \approx \gamma_{0} e^{\beta(|\bm{\delta}'_{i}|a^{-1}_{0}-1)}$ by considering changes of the distance between neighboring $2p_z$ atomic orbitals. Here, we set $\gamma_{0}$ to be the hopping energy for unstrained graphene and chose $\beta$ to be 3.37~\cite{ribeiro2009strained,PhysRevLett.103.046801}. The displacement vectors from the initial to neighboring sites are described as $\bm{\delta}'_{1}=(1/2-\sqrt{3}\zeta/2,-\sqrt{3}/2+\zeta/2)a$, $\bm{\delta}'_{2}=(1/2+\sqrt{3}\zeta/2,\sqrt{3}/2+\zeta/2)a$, and $\bm{\delta}'_{3}=(-1,-\zeta)a$, respectively, where $\zeta$ and $a$ are the shear-strain parameter and lattice constant of unstrained graphene.

HHG in shear-strained graphene is formulated by an extended theoretical framework based on the previous ones~\cite{Tamaya2016,Tamaya2016PRBR,Tamaya2019,Tamaya2021,Yoshikawa2017,xia2021high}. {The theoretical framework employed here is based on the tight-binding model that is known to provide quantitatively reasonable results for the band structures of shear-strained graphene \cite{cocco2010gap,si2016strain,li2010strain}.} We suppose the vector potential of the incident THz light to be $\bm{A}(t)=A_{0}f(t) \cos(\omega_{0} t) \, {\bm n}$, where $f(t)$ is the envelope function of the incident pulse, ${\bm n} = (\cos\theta, \sin\theta)$ is a unit vector pointing in the direction of the electric field, and $\theta$ is the tilt angle measured from the armchair axis (see Fig.~\ref{fig:config1}~(a)). We chose $f(t)=(t-t_{0})^2/\tau^2$ to be the envelope function, where $t_{0}=24 \pi/\omega_{0}$ and $\tau =4\pi/\omega_{0}$. The tight-binding Hamiltonian on a distorted honeycomb lattice in momentum space can be expressed as (see Supplementary Material: Section I)
\begin{align}
&H =\sum_{\bm{k}} ( e_{\bm k}^\dagger \ h_{-{\bm k}} )
\begin{pmatrix}
\xi({\bm k},t) & \eta({\bm k},t) \\
\eta({\bm k},t)^* & -\xi({\bm k},t) 
\end{pmatrix}\begin{pmatrix}
e_{\bm k} \\ h_{-{\bm k}}^\dagger
\end{pmatrix}, 
\label{Hamiltinian0}
\end{align}
where $e_{\bm{k}}$ ($h_{\bm{k}}$) is the annihilation operator of electrons (holes) with wavenumber $\bm{k}$, ${\xi}(\bm{k},t) = \left|f(\bm{k})\right| + \hbar \, {\rm Re} \, \left[\Omega^{\ast}_{R}(\bm{k},t) e^{i\theta_{\bm k}} \right]$, ${\eta}(\bm{k},t) = -\hbar \, {\rm Im} \, \left[\Omega^{\ast}_{R}(\bm{k},t) e^{i\theta_{\bm k}} \right]$, $f(\bm{k})=\sum_{i=1}^3 \gamma_{i}e^{i\bm{k}\cdot\bm{\delta'}_{i}}=|f(\bm{k})|e^{i\theta_{\bm{k}}}$, and $\Omega_{R}(\bm{k},t)={\bm d}({\bm k}) \cdot {\bm A}(t)$ is the Rabi frequency defined from the ${\bm k}$-dependent dipole vector ${\bm d}({\bm k})=\left(d^{\bm{k}}_{x},d^{\bm{k}}_{y}\right)$ (the explicit form of the dipole moment ${\bm d}({\bm k})$ is given in the Supplementary Material: Section I). We solved the time-dependent equations for polarizations and carrier densities derived from the Hamiltonian, and evaluated the HHG intensity spectra by performing a Fourier transformation on the generated current~\footnote{{Our theoretical framework is consistent with the previous ones \cite{malic2011microscopic,stroucken2011optical} in the limit of $\zeta=0$.}
To confirm the validity of this framework, we {also} checked that the exotic characteristics of HHG identified in the past experiment were reproduced (see Supplementary Material: section II).}. In this work, we only focused on HHG emitted along the major axis (parallel to ${\bm{A}}(t)$) and ignored the perpendicular component~\cite{Tamaya2021} {due to its complex behaviour as a function of the incident-light polarization angle and THz frequency}. 

Fig.~\ref{fig:config1} shows the 3D plot (upper figures) and contour plot (projection figures) of the band structure for three different shear strain parameters, (b1) $\zeta=0$, (b2) $\zeta=0.12$, and (b3) $\zeta=0.2$. For unstrained graphene ($\zeta=0$), we can see that Dirac cones exist at the $K^{\pm}$ points indicated by the white circles in Fig.~\ref{fig:config1}~(b1). These two Dirac cones gradually get closer to each other as the shear strain parameter increases (see Fig.~\ref{fig:config1}~(b2)), and consequently, they merge and an energy gap appears (see Fig.~\ref{fig:config1}~(b3)). In our tight-binding model, the critical strength of the shear strain for a band gap to form is estimated as $\zeta_{c} \approx 0.165$, which is consistent with previous works~\cite{PhysRevB.81.241412,sahalianov2019straintronics}.

The shear-strain dependences of the third and fifth harmonics are shown in Fig.~\ref{fig:config2}~(a1) and (b1) for an incident-light frequency of $\omega_{0}=90\, \rm{THz}$. The red and green lines indicate the dependences for the different incident-light polarization angles, $\theta = 120^{\circ}$ and $\theta = 30^{\circ}$, respectively. These two angles are regarded as the specific ones because they can yield the maximum and minimum intensities of the emitted harmonics (see insets in Fig.~\ref{fig:config2}~(a1) and (b1)). Fig.~\ref{fig:config2}~(a1) and (b1) show that, for $\theta = 120^{\circ}$, the third-order (fifth-order) harmonics are largely enhanced by around one order of magnitude at $\zeta \approx 0.13$ and $0.18$, while they are quenched rapidly by around third (eight) orders of magnitude at $0.2<\zeta<0.27$ (red line). The plots for $\theta = 30^{\circ}$ do not show this nonmonotonic behaviour; they only show an approximately monotonic decrease with increasing shear strain. Fig.~\ref{fig:config2}~(a2) and (b2) show the shear-strain dependence of the third- and fifth-order harmonics for the different incident-light frequencies, $\omega_0 = 30\,{\rm THz}$ (blue lines), $60\,{\rm THz}$ (orange lines), $90\,{\rm THz}$ (red lines), and $120\,{\rm THz}$ (green lines). Here, the interval between the two peaks grows with increasing incident-light frequency. This result suggests that HHG should be enhanced when the photon energy of the incident light coincides with a characteristic energy in the electronic structure.

\begin{figure}[tb]
\begin{center}
\includegraphics[width=8.9cm]{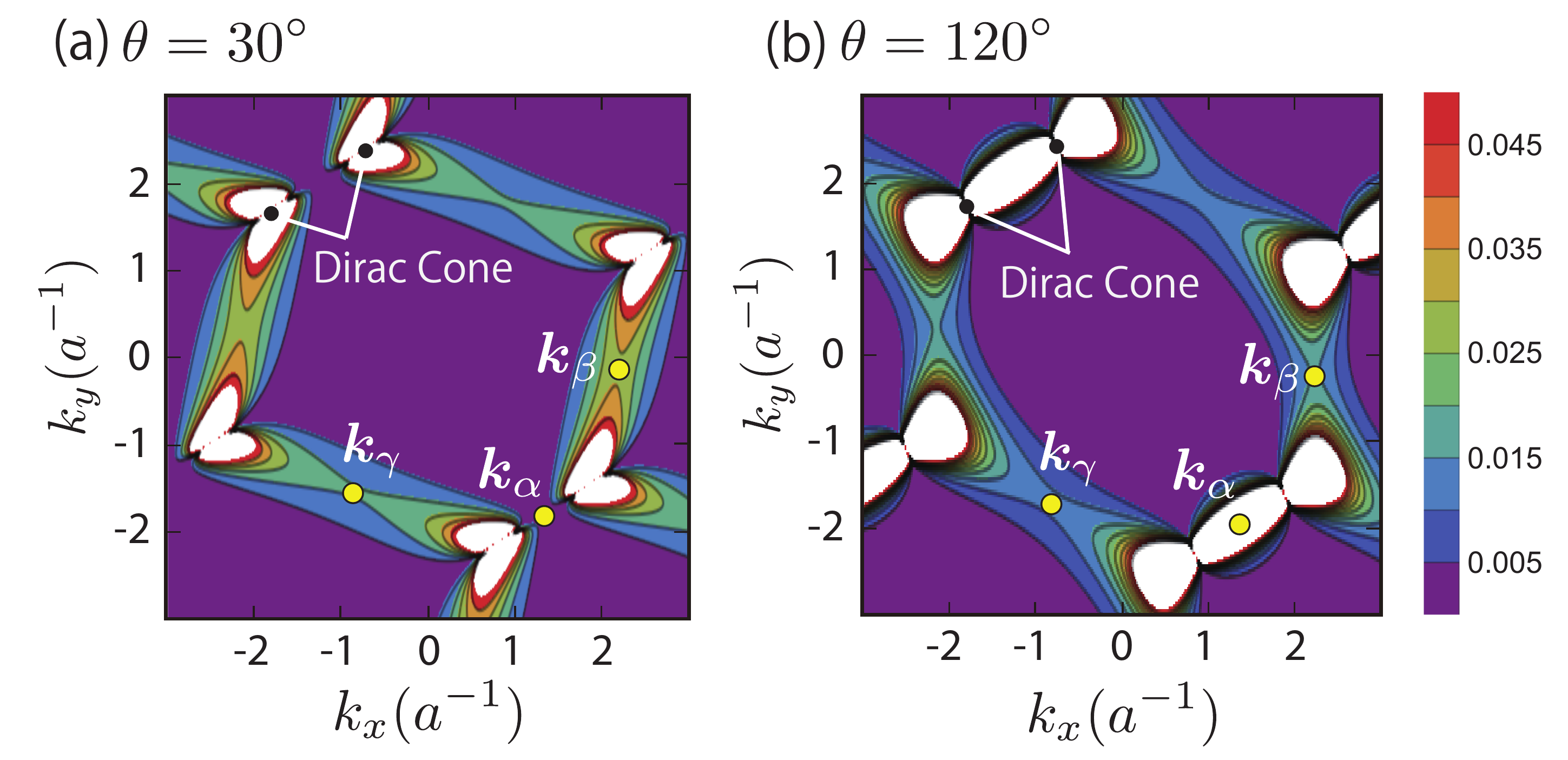}
\caption{(Color online) Contour plots of distribution of carrier density in $\bm{k}$-space after photo-irradiation. The incident-light polarization angle is (a) $\theta=30^{\circ}$ and (b) $\theta=120^{\circ}$. The shear-strain parameter and incident-light frequency are $\zeta=0.12$ and $\omega_{0}=90\rm{THz}$.
\label{fig:config4}}
\end{center}
\end{figure}

Strong incident-light polarization angle dependence is another remarkable feature of HHG in strained graphene (see Fig.~\ref{fig:config2}~(a1) and (b1)). This feature reflects the fact that the distribution of the excited carriers is sensitive to the polarization angle of the incident light. Fig.~\ref{fig:config4}~(a) and (b) show the distributions of the carrier density in $\bm{k}$-space after photo-irradiation for $\theta=30^{\circ}$ and $\theta=120^{\circ}$. Here, we set the shear-strain parameter and the incident-light frequency as $\zeta=0.12$ and $\omega_0 = 90\, \rm{THz}$, respectively. In these figures, the positions of the three saddle points of the band dispersion, ${\bm k}_\alpha$, ${\bm k}_\beta$, and ${\bm k}_\gamma$, are indicated by yellow points, and those of the two Dirac cones are denoted by black points. We find that the {carrier} density near the saddle point ${\bm k}_\alpha$ is suppressed for $\theta=30^{\circ}$ while it is enhanced for $\theta=120^{\circ}$. This fact indicates that the polarization angle of the incident light mainly affects the carrier excitation near the saddle point ${\bm k}_\alpha$.

\begin{figure*}[bt]
\begin{center}
\includegraphics[width=18cm]{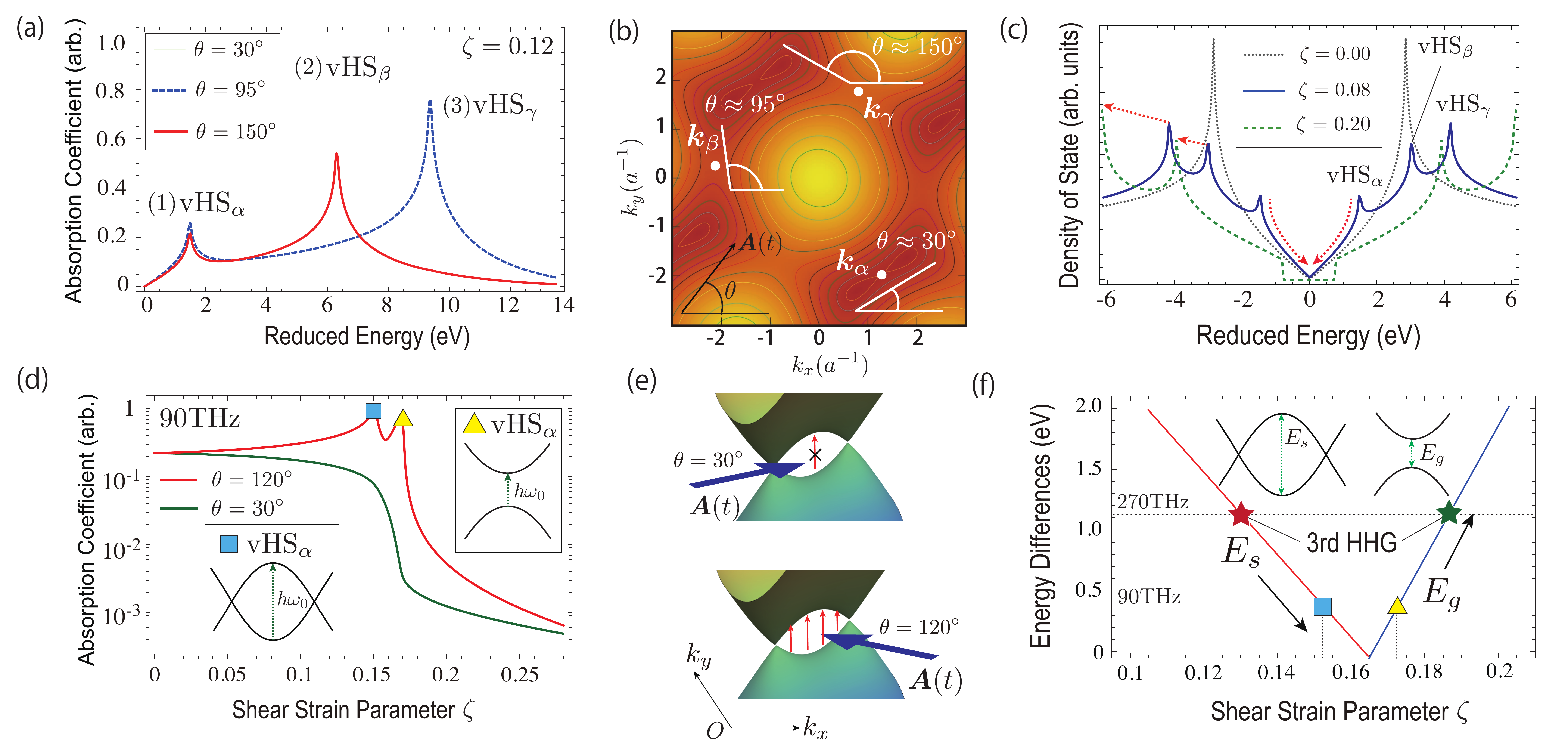}
\caption{(Color online) (a): Absorption coefficients of graphene for $\zeta=0.12$. The green, blue, and red lines show those for different incident-light polarization angles, $\theta=30^{\circ}$, $\theta=95^{\circ}$, and $\theta=120^{\circ}$, respectively. Each peak corresponds to different ${\rm{vHSs}}$, induced by the saddle points at $\bm{k}_{\alpha}$, $\bm{k}_{\beta}$, and $\bm{k}_{\gamma}$ shown in (b). (b): Contour plot of the energy band in graphene for $\zeta=0.12$. Here, ${\bm k}_{\alpha}$, ${\bm k}_{\beta}$, and ${\bm k}_{\gamma}$ indicate the positions of the saddle points in {$\bm{k}$-space}, each of which induces ${\rm vHS}_\alpha$, ${\rm vHS}_\beta$, and ${\rm vHS}_\gamma$, respectively. The azimuth angle toward the band minima (the Dirac cones) measured from the saddle points are indicated by the white lines. (c): DOS of graphene for the different shear strain parameters, $\zeta=0$ (black dotted line), $\zeta=0.12$ (blue line), and $\zeta=0.2$ (green dashed line). The red lines indicate that $\rm{vHS}$ shifts with increasing shear strain. (d): Absorption coefficients of graphene as a function of the shear strain parameter $\zeta$. The red and blue lines indicate those for the incident-light polarization angles $\theta=120^{\circ}$ and $\theta=30^{\circ}$. The insets show the resonant mechanisms corresponding to the double peaks of the red line (blue square and yellow triangles). (e): Schematic figure of photon absorption near $\rm{vHS}_{\alpha}$ for different incident-light polarization angles $\theta=30^{\circ}$ (upper figure) and $\theta=120^{\circ}$ (lower figure). {(f):} Energy differences at $\rm{vHS}_{\alpha}$ ($E_{s}$) and the band-gap energy $E_{g}$ as a function of shear strain. The blue square and yellow triangle indicate the resonant points marked in Fig.~\ref{fig:config3}(d). }
\label{fig:config3}
\end{center}
\end{figure*}

To better understand these features, we will investigate the characteristics of THz light absorption in shear-strained graphene. On the basis of linear response theory~\cite{kubo1957statistical}, we can derive the absorption coefficient $\alpha(\omega)$ as (see Supplementary Material: Section III)
\begin{align}
&\alpha(\omega) \simeq \frac{\pi}{\hbar \omega}\sum_{\bm{k}}(\beta_{x}^{\bm{k}}\cos\theta+\beta_{y}^{\bm{k}}\sin\theta)^{2} \nonumber \\ 
&\hspace{15mm} \times [\delta(\omega-2\epsilon_{\bm{k}})-\delta(\omega+2\epsilon_{\bm{k}})],
\label{Absorption}
\end{align}
where $\beta^{\bm{k}}_{x}$ and $\beta^{\bm{k}}_{y}$ are defined as $\beta^{\bm{k}}_{i}=-c\hbar({\rm{Im}}[d^{\bm{k}}_{i}]\cos\theta_{\bm{k}}-{\rm{Re}}[d^{\bm{k}}_{i}]\sin\theta_{\bm{k}})$ ($i=x,y$). Fig.~\ref{fig:config3}~(a) shows the frequency dependence of the absorption coefficients for the three different incident-light polarization angles, $\theta=30^{\circ}$ (green line), $95^{\circ}$ (blue line), and $120^{\circ}$ (red line) in the case of $\zeta=0.12$. These plots show that each peak originates from a singularity in the density of state (DOS), i.e., a van Hove singularity (vHS), which is induced by the saddle points of the band dispersion in $\bm{k}$-space. Fig.~\ref{fig:config3}~(b) is a contour plot of the band dispersion $\epsilon_{\bm k}=|f({ \bm k})|$ for $\zeta = 0.12$ as well as its saddle point. The carrier excitations near the three saddle points, labeled ${\bm k}_\alpha$, ${\bm k}_\beta$, and ${\bm k}_\gamma$, induce the peak structures denoted {as} ${\rm vHS}_{\alpha}$, ${\rm vHS}_{\beta}$, and ${\rm vHS}_{\gamma}$ in Fig.~\ref{fig:config3}~(a). Fig.~\ref{fig:config3}~(b) also shows that each saddle point has a different angle to the band minima (the Dirac points) that can be estimated as $\theta\approx 30^{\circ}$, $\theta \approx 95^{\circ}$, and $\theta \approx 150^{\circ}$ (see the white lines in Fig.~\ref{fig:config3}~(b)). We can identify that the peaks observed in Fig.~\ref{fig:config3}~(a) disappear when the direction of the incident light matches these angles, while the peak height in the absorption coefficient becomes a maximum when the incident light is perpendicular to them (see Supplementary Material: Section IV). Thus, the vHS peaks in the absorption coefficient significantly depend on the relation between the polarization angle of the incident light and the band structure. 
The position of the vHSs in DOS changes depending on the strength of the shear strain. Fig.~\ref{fig:config3}~(c) plots the DOS of graphene for $\zeta=0$ (black dotted line), $\zeta=0.08$ (blue line), and $\zeta=0.2$ (green dashed line). In the absence of shear strain ($\zeta=0$), only one vHS can be seen in the positive (negative) energy region. Applying shear strain splits it into three peaks (in each region) by breaking the energy degeneracy of the vHSs. The lowest-energy peak, labeled ${\rm vHS}_\alpha$, moves toward zero energy with increasing $\zeta$ (the blue line and the red curved arrow), and consequently, it reaches zero and the band gap starts to form (the green line). At the same time, the other two peaks, denoted by ${\rm vHS}_{\beta}$ and ${\rm vHS}_{\gamma}$, monotonically move toward the high-energy region (the red straight arrows). These numerical results imply that the properties of HHG in the THz regime ($\hbar \omega_0 \ll 1 \, {\rm eV}$) would be governed by the behavior of ${\rm vHS}_{\alpha}$.

Next, we focus on the absorption coefficient in the THz regime. Fig.~\ref{fig:config3}~(d) plots the shear-strain dependence of the absorption coefficient for $\theta=120^{\circ}$ (red line) and $\theta=30^{\circ}$ (green line) in the case of $\omega_{0}=90\,\rm{THz}$. This figure indicates that the absorption coefficient has two peaks for $\theta=120^{\circ}$, whereas the plot for $\theta = 30^{\circ}$ only shows a monotonic decrease. These tendencies are very similar to those of the HHG in Fig.~\ref{fig:config2}(a1) and (b1). As discussed so far, the properties of these peaks can be understood in terms of the relation between the incident-light polarization angle and $\rm{vHS}_{\alpha}$ (see Fig.~\ref{fig:config3}~(e)). Note that this angular dependence comes from the  dependence of the dipole moment $\bm{d}_{i}^{\bm{k}}$ on $\bm{k}$ through the coefficient $\beta_{x}^{\bm{k}}\cos\theta+\beta_{y}^{\bm{k}}\sin\theta$ in Eq.~(\ref{Absorption}). 
The electron-hole excitation energy at $\rm{vHS}_{\alpha}$ ($E_{s}$) and the band gap ($E_g$) are plotted as a function of the shear strain parameter $\zeta$ in Fig.~\ref{fig:config3}~(f). 
Comparing Fig.~\ref{fig:config3}~(d) and (f), we can conclude that the first peak in Fig.~\ref{fig:config3}~(d) (the blue square) should appear when the incident-light energy matches $E_{s}$ (see the lower inset), while the second peak (the yellow triangle) should appear when it matches $E_g$ (see the upper inset). This analysis enables us to conclude that the two HHG peaks in Fig.~\ref{fig:config2}~(a1) and (b1) arise from resonant conditions, $3\hbar \omega_{0}=E_{s}$ and $3\hbar \omega_{0}=E_{g}$, as indicated by the red and green star marks in Fig.~\ref{fig:config3}~(f). Note that the orientation of the dipole moment does not change much between before and after the energy gap forming. Thus, we can conclude that the significant enhancement in HHG is caused by the resonant mechanism at $\bm{k}_{\alpha}$ that changes greatly depending on the polarization angle of the incident light.

{So far, we have demonstrated that application of shear strain on graphene enables a significant control of HHG over several orders of magnitude. Note that it is fundamentally different from HHG modifications via control of carrier-envelope phase or chirp. The former controls material responses directly, while the latter use controlled incident light via some other methods. Therefore, our method provides a unique way for modulating or switching HHG of THz waves with a high dynamic range, which should be necessary in near-future high-speed wireless communication. A single-layer graphene may simultaneously solve problems of cost, mass, and volume of materials for an application of THz technology.}

In conclusion, we theoretically investigated the linear and nonlinear optical responses of graphene distorted by shear strain. Using the tight-binding model, we clarified that tuning both the shear strain and the polarization angle of the incident THz light enables a significant enhancement or quenching of HHG. By using linear response theory, we showed that the HHG enhancement came from the resonances at which the incident-light photon energy matches the saddle-point energy or the band gap. In addition, the dipole moment near the saddle point (or the band gap) plays an important role in the polarization angle dependence of this enhancement. Strong quenching of HHG was induced by rapid growth of the band-gap energy with increasing shear strain. Our findings suggest the possibility of using shear strain to control HHG in graphene and pave the way for optical nanotechnology of single-layer materials.

\vspace{2mm}
The authors acknowledge the support from the Japan Society for the Promotion of Science (JSPS KAKENHI Grants No. JP19K14624 and No. JP20K03831) and JST PRESTO (JPMJPR2107).

\appendix

\bibliography{GrapheneHHGref.bib}
\end{document}


\title{Supplemental material for Shear-Strain Controlled High-Harmonic Generation in Graphene}

\author{Tomohiro Tamaya}
\email{tamaya@g.ecc.u-tokyo.ac.jp}
\affiliation{Institute for Solid State Physics, University of Tokyo, Kashiwa, 277-8581, Japan}
\affiliation{JST, PRESTO, 4-1-8 Honcho, Kawaguchi, Saitama, 332-0012, Japan}

\author{Hidefumi Akiyama}
\affiliation{Institute for Solid State Physics, University of Tokyo, Kashiwa, 277-8581, Japan}

\author{Takeo Kato}
\affiliation{Institute for Solid State Physics, University of Tokyo, Kashiwa, 277-8581, Japan}
\date{\today}

\maketitle

\date{\today}

\maketitle 

\section{Detailed theoretical formulation}
\label{app:sec1}
In this section, we provide detailed information on the theoretical framework in the main text. In Sec.~\ref{app:subsec1}, we introduce the Hamiltonian of shear-strained graphene under terahertz (THz) light. In Sec.~\ref{app:subsec2}, we derive the time-evolution equations for the polarizations and carrier densities by utilizing the Hamiltonian derived in Sec.~\ref{app:subsec1} and define the current generated under shear strain. Note that our formulation is an extension of the theoretical framework employed in our previous works~\cite{Tamaya2016,Tamaya2016PRBR,Tamaya2019,Tamaya2021,Yoshikawa2017,xia2021high} and is consistent with other work as well~\cite{malic2011microscopic,stroucken2011optical}.

\subsection{Derivation of the Hamiltonian}
\label{app:subsec1}
\begin{figure}[tb]
\centering
\includegraphics[width=16cm]{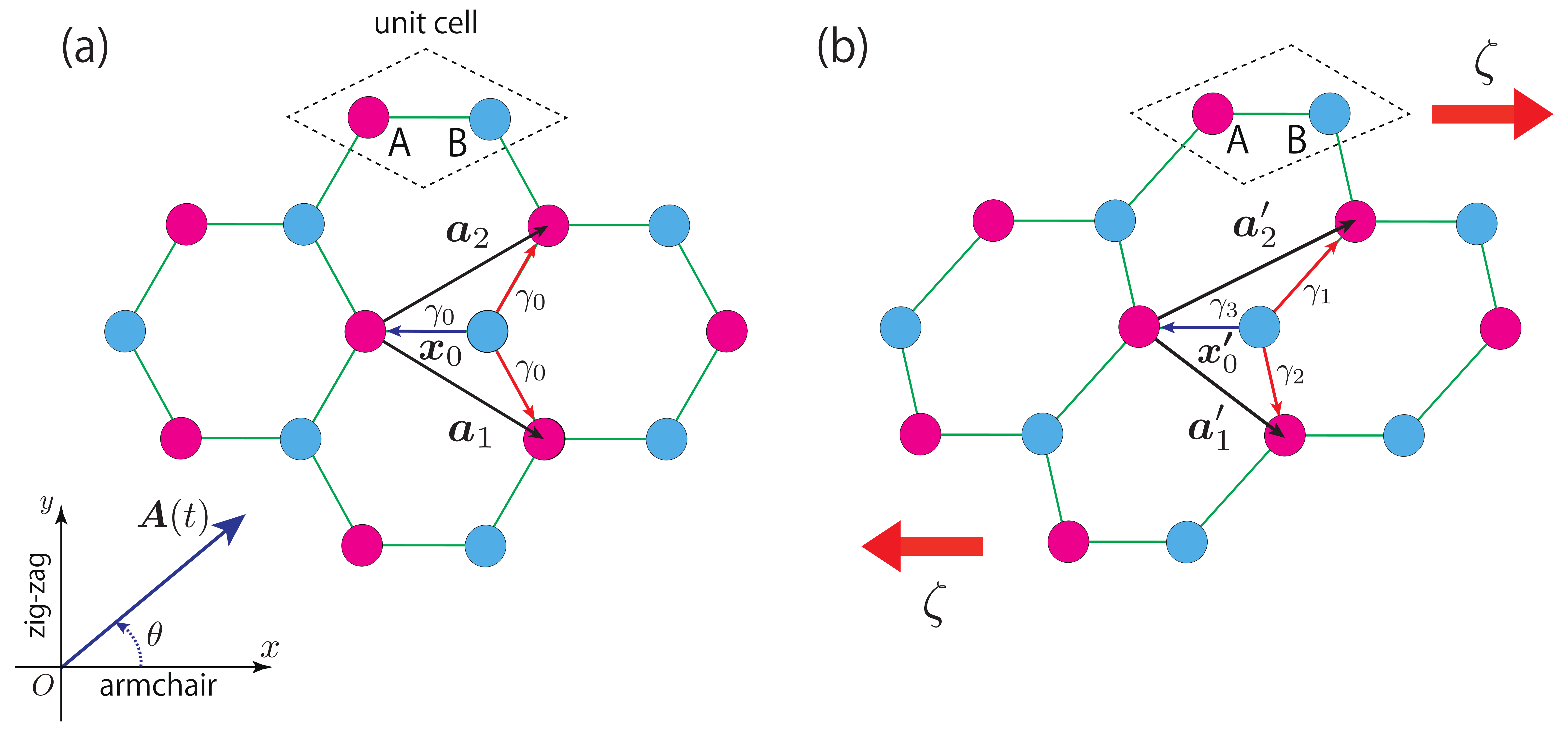}
\caption{(Color online) Schematic diagram of lattice structure of graphene (a) in the absence of shear strain and (b) under shear strain. The dotted rhomboid area indicates the unit cell including two carbon atoms, A and B. Under shear strain, the basic lattice vectors, $\bm{a}_{1}$ and $\bm{a}_{2}$, are modified into $\bm{a}_{1}'$ and $\bm{a}_{2}'$. Correspondingly, the hopping energies that take a common value $\gamma_0$ in the absence of shear strain are modified into $\gamma_i$ ($i=1,2,3$). The inset of (a) indicates the direction of the vector potential, which is parallel to the electric field of the incident light.}
\label{fig:configS1}
\end{figure}

We start with the well-known form of the Hamiltonian,
\begin{align}
\hat{H}=\frac{1}{2m_0}\left(\bm{p}-\frac{e}{c}\bm{A}(t)\right)^2+\sum_{i}V(\bm{x}-\bm{R}_{i}), 
\end{align}
where $m_{0}$ is the electron mass, $e$ ($<0$) is the electron charge, $\bm{p}$ is the momentum of the bare electron, $c$ is the velocity of light, $\bm{A}(t)$ is the vector potential of the incident THz light, and $V(\bm{x}-\bm{R}_{i})$ is the core potential of an atom located at $\bm{R}_{i}$. We employ the Coulomb gauge~\cite{haug2009quantum}. By expanding this Hamiltonian, we obtain
\begin{align}
\hat{H}&=\hat{H}_{0}+\hat{H}_{I}, \nonumber \\
\hat{H}_{0}&= \frac{\bm{p}^2}{2m_{0}}+\sum_{i}V(\bm{x}-\bm{R}_{i}), \nonumber \\
\hat{H}_{I}&=-\frac{e}{m_{0}c}\bm{A}(t)\cdot\bm{p}+\frac{e^2 \bm{A}^2(t)}{2 m_{0} c^2}. \nonumber
\end{align}
The second term in $\hat{H}_{I}$ can be eliminated by performing a unitary transformation $U^{-1}_1 \hat{H}_{I} U_1$, where 
\begin{align}
U_1 = \exp\left[\frac{ie^2}{2m_0c^2\hbar^2}\int^t_0 dt' \, {\bm A}^2(t') \right].
\end{align}
Thus, we obtain the total Hamiltonian of the system in the form,
\begin{align}
\hat{H} &= \hat{H}_0+\hat{H}_{I}, \nonumber \\
\hat{H}_0&=\frac{\bm{p}^2}{2m_0}+\sum_{i}V(\bm{x}-\bm{R}_{i}), \label{Hamiltonian:00} \\
\hat{H}_{I}&=-\frac{e}{m_{0}c}{\bm{A}(t)} \cdot {\bm{p}}. \label{Hamiltonian:01} 
\end{align}
Here, we assume that the incident THz light is linearly polarized. The vector potential of the THz light is defined as $\bm{A}(t)=A_{0}f(t) \cos(\omega_{0} t) \, {\bm n}$, where $f(t)$ is the envelope function of the incident pulse, ${\bm n} = (\cos\theta, \sin\theta)$ is a unit vector pointing in the direction of the electric field, and $\theta$ is the tilt angle with respect to the armchair axis (see Fig.~\ref{fig:configS1}~(a)). In our calculation, we chose the envelope function to be $f(t)=(t-t_{0})^2/\tau^2$, where $t_{0}=24 \pi/\omega_{0}$ and $\tau =4\pi/\omega_{0}$.

The lattice structure of graphene is defined as follows. As shown in Fig.~\ref{fig:configS1}~(a), the unit cell includes two carbon atoms, A and B, as shown by the dotted rhomboid area, and the lattice vectors of unstrained graphene are set as $\bm{a}_{1}=(3/2,-\sqrt{3}/2)a_{0}$ and $\bm{a}_{2}=( 3/2,\sqrt{3}/2)a_{0}$, where $a_{0}$ is the lattice constant and the $x$- and $y$-axis are the armchair and zigzag directions, respectively. Accordingly, the reciprocal lattice vectors are $\bm{b}_{1}=(2 \pi/3)(1,\sqrt{3})a^{-1}_{0}$ and $\bm{b}_{2}=(2 \pi/3)(1,-\sqrt{3})a^{-1}_{0}$.

Here, we introduce the strain tensor $\xi$ in the form, 
\begin{align}
\xi=\left( \begin{array}{cc} 
0 & \zeta \\ \zeta & 0
\end{array} \right),
\end{align}
where $\zeta$ is the strength of in-plane shear strain indicated by the red arrows in Fig.~\ref{fig:configS1}~(b). According to the Cauchy-Born rule, the lattice vectors ${\bm a}=(a_1,a_2)$ are modified by the shear strain into $a'_{i}=a_{i}+\sum_{j}{\xi}_{ij}a_{j}$ ($i,j=1,2$)\cite{cocco2010gap}. Therefore, the lattice vectors of graphene are modified into
\begin{align}
 \bm{a}_{1}' &= \left(\frac{3}{2} -\frac{\sqrt{3}}{2}\zeta,-\frac{\sqrt{3}}{2}+\frac{3}{2}\zeta\right)a_{0},\\
 \bm{a}_{2}' &= \left(\frac{3}{2} +\frac{\sqrt{3}}{2}\zeta,\frac{\sqrt{3}}{2}+\frac{3}{2}\zeta\right)a_{0}.
\end{align} 
(See Fig.~\ref{fig:configS1}~(b).) The corresponding reciprocal lattice vectors are 
\begin{align}
\bm{b}_{1}'&=\left(\frac{2\pi\left(-1+\sqrt{3}\zeta\right)}{3\left(\zeta^2-1\right)},-\frac{2\pi\left(-\zeta+\sqrt{3}\right)}{3\left(\zeta^2-1\right)}\right) {a_0^{-1}} , \\
\bm{b}_{2}'&=\left(-\frac{2\pi\left(1+\sqrt{3}\zeta\right)}{3\left(\zeta^2-1\right)},\frac{2\pi\left(\zeta+\sqrt{3}\right)}{3\left(\zeta^2-1\right)}\right) {a_0^{-1}} .
\end{align}

The second-quantized Hamiltonian of graphene is
\begin{align}
H_0=\sum_{n\nu n'\nu'\bm{k}\bm{k'}}\braket{\bm{k}n\nu|\hat{H}_{0}|\bm{k'}n'\nu'}c^{\dagger}_{n\nu\bm{k}}c_{n'\nu'\bm{k'}}, 
\end{align}
where $c_{n \nu \bm{k}}$ is the annihilation operator of electrons with wavenumber $\bm{k}$ that are bound to an atom $\nu$ (=A or B) located at $\bm{R}_{n\nu}$. Here, we consider only the $2p_z$ orbital of graphene, whose wavefunctions are given as
\begin{align}
\phi_{2p_{z}}(\bm{x})=\frac{z}{4\sqrt{2\pi d^5}}\exp\left[-\frac{1}{2d}\sqrt{x^2+y^2+z^2}\right],
\end{align}
where $d = a_{B}/Z_{\rm eff}$ is a parameter controlling the effective spread of the $2p_z$ orbital. We set $d$ to 2.461 \AA. The tight-binding wavefunction is expressed as 
\begin{align}
\braket{\bm{x}|n\nu\bm{k}}=\sum_{n}e^{i\bm{k}\cdot\bm{R}_{n\nu}}\phi_{2p_{z}}(\bm{x}-\bm{R}_n).
\end{align}
For a tight-binding model on a two-dimensional honeycomb lattice with nearest-neighbor hopping, the Hamiltonian is given by
\begin{align}
H_{0} &=\sum_{\bm{k}}f(\bm{k})a^{\dagger}_{\bm{k}}b_{\bm{k}}+{\rm h.c.},
\label{Hamiltonian:03} \\
f(\bm{k})&=\sum_{i=1}^3 \gamma_{i}e^{i\bm{k}\cdot\bm{\delta'}_{i}},
\label{fkdef}
\end{align}
where $a_{\bm{k}}$ ($b_{\bm{k}}$) is the annihilation operator of electrons the wavenumber $\bm{k}$ that are on sublattice A (B), the three replacement vectors $\bm{\delta}'_{i}$ ($i=1,2,3$) from an A site to its nearest-neighbor B sites are given as $\bm{\delta}_{1}'=\bm{a}'_{1}+\bm{x}_{0}'$, $\bm{\delta}_{2}'=\bm{a}'_{2}+\bm{x}_{0}'$, and $\bm{\delta}_{3}'=\bm{x}_{0}'$, (see Fig.~\ref{fig:configS1}~(b)), and the nearest-neighbor hopping energies are defined as
\begin{align}
 \gamma_{i}=\int d\bm{x}'\phi_{2p_{z}}(\bm{x}')\hat{H}_{0}\phi_{2p_{z}}(\bm{x}'-\bm{\delta}_{i}'), \quad (i=1,2,3).
\end{align}
In accordance with the previous work~\cite{pereira2009strain,ribeiro2009strained}, $\gamma_{i}$ is approximately expressed as $\gamma_{i} \approx \gamma_{0} e^{\beta(|\bm{\delta}'_{i}|a^{-1}_{0}-1)}$, where $\gamma_{0}\approx2.78$ eV is the hopping energy for unstrained graphene and $\beta \approx 3.37$.

Using the tight-binding approximation, the second-quantized Hamiltonian for the light-matter interaction is obtained as
\begin{align}
H_{I}
&={\sum_{n\bm{k}\bm{k}'}\braket{\bm{k}nA|\hat{H}_{I}|\bm{k}'n B}a^{\dagger}_{\bm{k}}b_{\bm{k}'}+{\rm h.c.}} \nonumber \\
&=\sum_{n\bm{k}\bm{k}'i}e^{i(\bm{k}-\bm{k}')\cdot \bm{R}_{n}}e^{i\bm{k}\cdot \bm{\delta}'_{i}}a^{\dagger}_{\bm{k}}b_{\bm{k}'}\int d\bm{x'}\phi_{2p_z}(\bm{x}')\left( -\frac{e}{m_{0}c}\bm{A}(t)\cdot \bm{p}\right)\phi_{2p_z}(\bm{x}'-\bm{\delta}_{i}')+{\rm h.c.} \nonumber \\
&=\hbar \sum_{\bm{k}}\left[\Omega_{R}(\bm{k},t)a^{\dagger}_{\bm{k}}b_{\bm{k}}+\Omega^{\ast}_{R}(\bm{k},t)b^{\dagger}_{\bm{k}}a_{\bm{k}} \right],
\end{align}
where the Rabi frequency $\Omega_{R}(\bm{k},t)$ is defined as
\begin{align}
\Omega_{R}(\bm{k},t)&=-\frac{e\hbar}{m_{0}c} \sum_{i}e^{i \bm{k}\cdot \bm{\delta}'_{i}}\int d\bm{x} \phi_{2p_{z}}(\bm{x}')\bm{A}(t)\cdot \bm{p} \phi_{2p_{z}}(\bm{x}'-\bm{\delta}'_{i}) \nonumber \\
&\equiv d_{x}({\bm{k}}){A}_{x}(t)+d_{y}({\bm{k}}){A}_{y}(t). 
\label{Rabidef}
\end{align}
Here, $d_{x}(\bm{k})$ and $d_{y}(\bm{k})$ are the $x$- and $y$ components of the dipole moments of the shear-strained graphene. For convenience, the strength of the incident THz light in our calculation is set to satisfy $|d_{x}({\bm{k}_{0}}){A}_{0}|/\omega_{0} = 1.92$ in unstrained graphene in order to compare numerical results for different incident-light frequencies, where ${{\bm k}_{0}}$ is the wavenumber that maximizes $d_{x}(\bm{k})$ in the Brillouin zone.

In summary, the total tight-binding Hamiltonian is 
\begin{align}
H&=H_0+H_{I}, \nonumber \\
H_{0}&=\sum_{\bm{k}}\left[f(\bm{k})a^{\dagger}_{\bm{k}}b_{\bm{k}}+f^{\ast}(\bm{k})b^{\dagger}_{\bm{k}}a_{\bm{k}} \right],
\label{H0sum} \\
H_{I}&=\hbar \sum_{\bm{k}}\left[\Omega_{R}(\bm{k},t)a^{\dagger}_{\bm{k}}b_{\bm{k}}+\Omega^{\ast}_{R}(\bm{k},t)b^{\dagger}_{\bm{k}}a_{\bm{k}} \right],
\label{HI}
\end{align}
where $f(\bm{k})$ and $\Omega_{R}(\bm{k},t)$ are given by Eqs.~(\ref{fkdef}) and (\ref{Rabidef}), respectively. The Hamiltonian is diagonalized by a unitary transformation, defined as
\begin{align}
a_{\bm{k}}&=\frac{1}{\sqrt{2}}[e_{\bm{k}}+i h^{\dagger}_{-\bm{k}}], \\
b_{\bm{k}}&=\frac{1}{\sqrt{2}}e^{i\theta_{\bm{k}}}
[e_{\bm{k}}-i h^{\dagger}_{-\bm{k}}] ,
\label{Unitary}
\end{align}
where $\theta_{\bm{k}}$ is the argument of $f(\bm{k})$, defined as $f(\bm{k})=\left|f(\bm{k})\right|e^{i\theta_{\bm{k}}}$. Substituting these expressions into Eqs.~(\ref{H0sum}) and (\ref{HI}) yields
\begin{align}
H&=H_{0}+H_{I}, \nonumber \\
H_0&=\sum_{\bm{k}} \left|f(\bm{k})\right| [e^{\dagger}_{\bm{k}}e_{\bm{k}}-h_{-\bm{k}}h^{\dagger}_{-\bm{k}}], \label{Hamiltonian001} \\
H_{I}&=\hbar \sum_{\bm{k}}\left[{\rm{Re}}\left[\Omega_{R}(\bm{k},t)\right]\cos{\theta_{\bm{k}}}+ {\rm{Im}}\left[\Omega_{R}(\bm{k},t)\right]\sin{\theta_{\bm{k}}}\right](e^{\dagger}_{\bm{k}}e_{\bm{k}}+h^{\dagger}_{-\bm{k}}h_{-\bm{k}}-1) \nonumber \\
&+\hbar \sum_{\bm{k}}\left[{\rm{Im}}\left[\Omega_{R}(\bm{k},t)\right]\cos{\theta_{\bm{k}}}-{\rm{Re}}\left[\Omega_{R}(\bm{k},t)\right]\sin{\theta_{\bm{k}}}\right](h_{-\bm{k}}e_{\bm{k}}+e^{\dagger}_{\bm{k}}h^{\dagger}_{-\bm{k}}) ,
\label{Hamiltonian002}
\end{align} 
where $\cos{\theta_{\bm{k}}}={\rm{Re}}\left[f(\bm{k})/|f(\bm{k})|\right]$ and $\sin{\theta_{\bm{k}}}={\rm{Im}}\left[f(\bm{k})/|f(\bm{k})|\right]$. We can rewrite the Hamiltonian in matrix form as 
\begin{align}
&H =\sum_{\bm{k}} ( e_{\bm k}^\dagger \ h_{-{\bm k}} )
\begin{pmatrix}
\xi({\bm k},t) & \eta({\bm k},t) \\
\eta({\bm k},t)^* & -\xi({\bm k},t) 
\end{pmatrix}
\begin{pmatrix}
e_{\bm k} \\ h_{-{\bm k}}^\dagger
\end{pmatrix}, \label{HH} 
\end{align}
where ${\xi}(\bm{k},t) = \left|f(\bm{k})\right| + \hbar \, {\rm Re} \, \left[\Omega^{\ast}_{R}(\bm{k},t) e^{i\theta_{\bm k}} \right]$ and ${\eta}(\bm{k},t) = -\hbar \, {\rm Im} \, \left[\Omega^{\ast}_{R}(\bm{k},t) e^{i\theta_{\bm k}} \right]$, respectively. This expression corresponds to Eq.~(1) in the main text.

\subsection{Derivation of the time-evolution equations}
\label{app:subsec2}
Here, we derive the time evolution equations for the polarization $P_{\bm{k}}=\braket{h_{-\bm{k}}e_{\bm{k}}}$ and the carrier densities $f^{\sigma}_{\bm{k}}=\braket{\sigma_{\bm{k}}^{\dagger}\sigma_{\bm{k}}}$ ($\sigma=e,h$). The total Hamiltonian Eq.(\ref{HH}) can be rewritten as
\begin{align}
&H=\sum_{\bm{k}}{\xi}(\bm{k}{,t})[e^{\dagger}_{\bm{k}}e_{\bm{k}} - h_{-\bm{k}}h^{\dagger}_{-\bm{k}}] \nonumber \\
&\hspace{10mm} +\hbar \sum_{\bm{k}}\Bigl[{\rm{Im}}\left[\Omega_{R}(\bm{k},t)\right]\cos{\theta_{\bm{k}}}-{\rm{Re}}\left[\Omega_{R}(\bm{k},t)\right]\sin{\theta_{\bm{k}}}\Bigr](h_{-\bm{k}}e_{\bm{k}}+e^{\dagger}_{\bm{k}}h^{\dagger}_{-\bm{k}}),
\label{Hamiltonian4} \\
& {\xi}(\bm{k},t)= \left|f(\bm{k})\right| + \hbar \Bigl[{\rm{Re}}\left[\Omega_{R}(\bm{k},t)\right]\cos{\theta_{\bm{k}}}+{\rm{Im}}\left[\Omega_{R}(\bm{k},t)\right]\sin{\theta_{\bm{k}}}\Bigr] .
\end{align}
As discussed in our previous works~\cite{Tamaya2016,Tamaya2016PRBR,Tamaya2019,Tamaya2021,Yoshikawa2017,xia2021high}, ${\xi}(\bm{k},t)$ is the field-modulated band dispersion that becomes more significant with increasing incident field intensity. Using this Hamiltonian, the equations of motion for $e_{\bm{k}}$ and $h_{-\bm{k}}$ can be derived as
\begin{align}
i\hbar \dot{e}_{\bm{k}}&= [e_{\bm{k}},H] \nonumber \\
&={\xi}(\bm{k}{,t})e_{\bm{k}}+\hbar\left[{\rm{Im}}\left[\Omega_{R}(\bm{k},t)\right]\cos{\theta_{\bm{k}}}-{\rm{Re}}\left[\Omega_{R}(\bm{k},t)\right]\sin{\theta_{\bm{k}}}\right]h^{\dagger}_{-\bm{k}}, 
\label{Kinetic_Equations1} \\
i\hbar \dot{h}_{-\bm{k}}&= [h_{-\bm{k}},H] \nonumber \\
&={\xi}(\bm{k}{,t})h_{-\bm{k}}+\hbar \left[{\rm{Im}}\left[\Omega_{R}(\bm{k},t)\right]\cos{\theta_{\bm{k}}}-{\rm{Re}}\left[\Omega_{R}(\bm{k},t)\right]\sin{\theta_{\bm{k}}}\right]e^{\dagger}_{\bm{k}}. 
\label{Kinetic_Equations2}
\end{align}
By using these equations, we can further derive equations for the time evolution of the polarization $P_{\bm{k}}=\braket{h_{-\bm{k}}e_{\bm{k}}}$ and the carrier densities $f^{\sigma}_{\bm{k}}=\braket{\sigma_{\bm{k}}^{\dagger}\sigma_{\bm{k}}}$ ($\sigma=e,h$) as
\begin{align}
i\hbar \dot{P}_{\bm{k}}&=2{\xi}(\bm{k}{,t})P_{\bm{k}}+\hbar \Bigl[{\rm{Im}}\left[\Omega_{R}(\bm{k},t)\right]\cos{\theta_{\bm{k}}}-{\rm{Re}}\left[\Omega_{R}(\bm{k},t)\right]\sin{\theta_{\bm{k}}}\Bigr](1-f^{e}_{\bm{k}}-f^{h}_{\bm{k}}), \\
i\hbar \dot{f}^{\sigma}_{\bm{k}}&=-2\Bigl[{\rm{Im}}\left[\Omega_{R}(\bm{k},t)\right]\cos{\theta_{\bm{k}}}-{\rm{Re}}\left[\Omega_{R}(\bm{k},t)\right]\sin{\theta_{\bm{k}}}\Bigr]{\rm{Im}}\left[P_{\bm{k}}\right].
\end{align} 
We assume that the system is initially in the ground state, i.e., $f^e_{\bm k} = f^h_{\bm k}=P_{\bm{k}}=0$. The solutions of these equations provide the time evolution of ${f}^{\sigma}_{\bm{k}}(t)$ and ${P}_{\bm{k}}(t)$. The generated current is defined as
\begin{align}
{\cal J}_i{(t)}=-c \Braket{\frac{\partial H_{I}}{\partial A_{i}(t)}}, \quad (i=x,y).
\label{CurrentDefinition}
\end{align}
By substituting the explicit form of $H_I$, we obtain
\begin{align}
{\cal J}_{x}{(t)} &= -c\hbar\sum_{\bm{k}} \biggl[\Bigl({{\rm{Re}}[d_{x}(\bm{k})]\cos {\theta_{\bm{k}}}+{\rm{Im}}}[d_{x}(\bm{k})]\sin {\theta_{\bm{k}}}\Bigr)(f^{e}_{\bm{k}}+f^{h}_{\bm{k}}-1) \nonumber \\
&\hspace{10mm} +\Bigl({{\rm{Im}}[d_{x}(\bm{k})]\cos {\theta_{\bm{k}}}-{\rm{Re}}}[d_{x}(\bm{k})]\sin {\theta_{\bm{k}}}\Bigr)(P_{\bm{k}}+P^{\dagger}_{\bm{k}})\biggr], \nonumber \\
{\cal J}_{y}{(t)} &= -c\hbar \sum_{\bm{k}} \biggl[\Bigl({{\rm{Re}}[d_{y}(\bm{k})]\cos {\theta_{\bm{k}}}+{\rm{Im}}}[d_{y}(\bm{k})]\sin {\theta_{\bm{k}}}\Bigr)(f^{e}_{\bm{k}}+f^{h}_{\bm{k}}-1) \nonumber \\
&\hspace{10mm} +\Bigl({{\rm{Im}}[d_{y}(\bm{k})]\cos {\theta_{\bm{k}}}-{\rm{Re}}}[d_{y}(\bm{k})]\sin {\theta_{\bm{k}}}\Bigr)(P_{\bm{k}}+P^{\dagger}_{\bm{k}})\biggr].
\label{Kinetic_Equations3}
\end{align}
The currents generated along to the major axis, which is the axis parallel to the vector potential of the incident light, are defined by ${\cal J}(t)={\cal J}_{x}(t)\cos\theta + {\cal J}_{y}(t)\sin\theta =\bm{{\cal J}}(t)\cdot \bm{n}$, where $\bm{{\cal J}}(t)=({\cal J}_{x}(t),{\cal J}_{y}(t))$ and $\bm{n}=(\cos\theta,\sin\theta)$. The HHG intensity spectra is calculated from the generated current as $I=\left|\omega {\cal J}(\omega) \right|^2$, where ${\cal J}(\omega)$ is the Fourier transform of ${\cal J}(t)$. Note that the emitted light may have also a component parallel to the minor axis, which is perpendicular to the major axis, in the shear-strained system \cite{Tamaya2021}.

\begin{figure}[tb]
\begin{center}
\includegraphics[width=16cm]{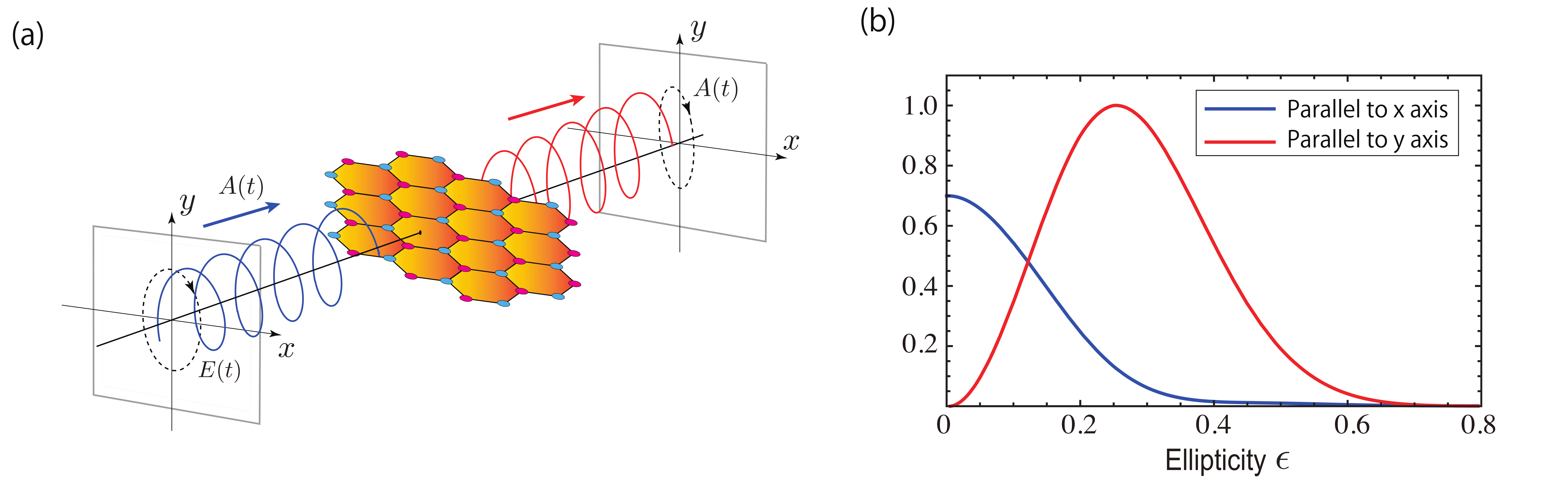}
\caption{(Color online) (a) Schematic diagram of HHG from unstrained graphene caused by elliptically polarized electric field. (b) Ellipticity dependence of 7th-order harmonics  emitted along to the major axis ($x$-axis, in blue) and along the minor axis ($y$-axis, in red).
\label{fig:configS3}}
\end{center}
\end{figure}

\section{Ellipticity dependence of high-harmonic generation in unstrained graphene}
It has been experimentally shown that high harmonics emitted parallel to the minor axis in graphene ($y$-axis in Fig.~\ref{fig:configS3}~(a)) show a maximum value at around $\epsilon \approx 0.3$~\cite{Yoshikawa2017}, where $\epsilon$ is the ellipticity of the incident electric field. This ellipticity dependence of HHG has been replicated by a simple theory of graphene that only focuses on the nearest Dirac cones and entirely ignores the $\bm{k}$-dependence of the Rabi frequency in Eq.~(\ref{Rabidef})~\cite{Tamaya2016,Tamaya2016PRBR,Yoshikawa2017}. Here, we check that the tight-binding model with the $\bm{k}$-dependent Rabi frequency, which is employed in our work, produces almost the same results. We set the vector potential as $\bm{A}(t)=A_{0}f(t) (\cos\omega_{0} t, \epsilon \sin\omega_{0} t)$ and calculate HHG in unstrained graphene as a function of the ellipticity $\epsilon$. Fig.~\ref{fig:configS3}~(b) shows the numerical results for 7th-order harmonics. As identified in the previous theory~\cite{Tamaya2016PRBR} and experiment~\cite{Yoshikawa2017}, the intensity of high harmonics parallel to the major axis (blue line) shows a Gaussian dependence on $\epsilon$, while that of harmonics parallel to the minor axis (red line) takes a maximum value for a finite ellipticity ($\epsilon \approx 0.3$). This means that the ellipticity dependence of HHG can be captured by the low-energy effective theory near the K and K' points.

\section{Absorption Coefficients of Graphene under Shear Strain}
\label{app:subsec3}
Here, we derive an expression of the absorption coefficient of graphene under shear strain within linear response theory. In Eq.~(\ref{Hamiltonian002}), the light-matter interaction Hamiltonian can be described as
\begin{align}
H_{I}&=\hbar \sum_{\bm{k}}\left[{\rm{Re}}\left[\Omega_{R}(\bm{k},t)\right]\cos{\theta_{\bm{k}}}+ {\rm{Im}}\left[\Omega_{R}(\bm{k},t)\right]\sin{\theta_{\bm{k}}}\right](e^{\dagger}_{\bm{k}}e_{\bm{k}}+h^{\dagger}_{-\bm{k}}h_{-\bm{k}}-1) \nonumber \\
&+\hbar \sum_{\bm{k}}\left[{\rm{Im}}\left[\Omega_{R}(\bm{k},t)\right]\cos{\theta_{\bm{k}}}-{\rm{Re}}\left[\Omega_{R}(\bm{k},t)\right]\sin{\theta_{\bm{k}}}\right](h_{-\bm{k}}e_{\bm{k}}+e^{\dagger}_{\bm{k}}h^{\dagger}_{-\bm{k}}),
\label{IHamiltonian21}
\end{align} 
where $\Omega_{R}(\bm{k},t)=d_{x}({\bm{k}}){A}_{x}(t)+d_{y}({\bm{k}}){A}_{y}(t)$. By the definitions (Eqs.~(\ref{CurrentDefinition}) and (\ref{Kinetic_Equations3})), the generated current is ${\cal J}_i(t) = \sum_{\bm k} \langle j_{{\bm k},i}\rangle$ ($i=x,y$), where $j_{{\bm k},i}$ is the current operator defined as
\begin{align}
j_{{\bm{k}},i} &= \alpha_i({\bm{k}})(e^{\dagger}_{\bm{k}}e_{\bm{k}}+h^{\dagger}_{-\bm{k}}h_{-\bm{k}}-1)+
\beta_i({\bm{k}})(h_{-\bm{k}}e_{\bm{k}}+e^{\dagger}_{\bm{k}}h^{\dagger}_{-\bm{k}}), \quad (i=x,y).
\label{DGCurrents}
\end{align} 
The coefficients, $\alpha_i({\bm k})$ and $\beta_i({\bm k})$ ($i=x,y$), are 
\begin{align}
&\alpha_{i}({\bm k})=-c\hbar\left[{\rm{Re}}[d_{i}({\bm k})]\cos\theta_{\bm{k}}+{\rm{Im}}[d_i({\bm{k}})]\sin\theta_{\bm{k}}\right], \\
&\beta_{i}({\bm k})=-c\hbar\left[{\rm{Im}}[d_i({\bm k})]\cos\theta_{\bm{k}}-{\rm{Re}}[d_i({\bm{k}})]\sin\theta_{\bm{k}}\right]. 
\label{AB}
\end{align}
In linear response theory, the optical conductivity $\sigma_{ij}(i,j=x,y)$ is defined by ${\cal J}_{i}(\omega)=\sigma_{ij}(\omega) E_j(\omega)$. From the Kubo formula~\cite{kubo1957statistical}, the optical conductivity can be written as
\begin{equation}
\sigma_{ij}(\omega)=\frac{1}{\hbar\omega}\sum_{\bm k} \int_{0}^{\infty} dt \langle [j^I_{\bm{k},i}(t),j^I_{\bm{k},j}(0)]\rangle_0 e^{i\omega t-\delta t},
\label{Kubo1}
\end{equation} 
where $j^I_{\bm{k},i}(t)=e^{iH_{0}t}j^{\bm{k}}_{i}e^{-iH_{0}t}$ is the current operator in the interaction picture and $\langle \cdots \rangle_0$ indicates the expectation value with respect to the ground state.

\begin{figure}[tb]
\begin{center}
\includegraphics[width=8cm]{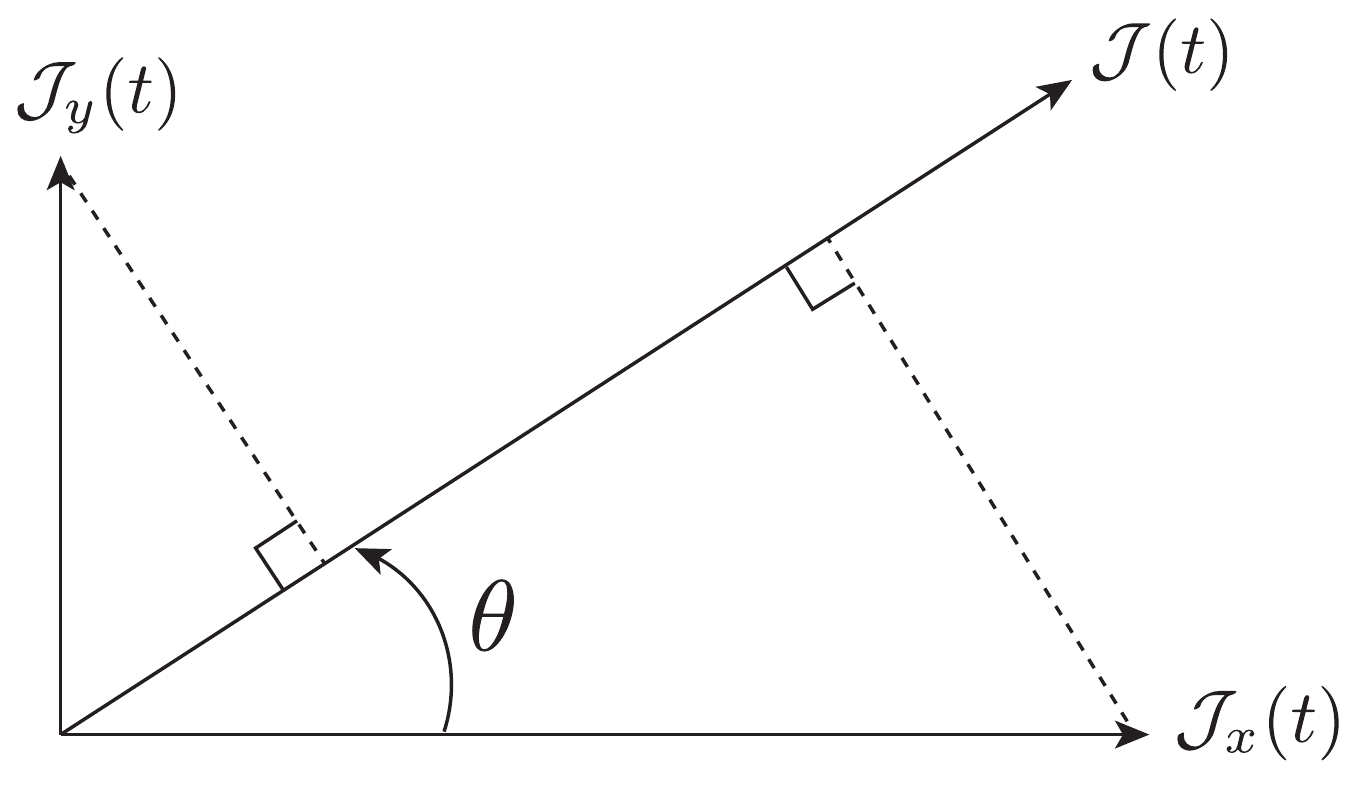}
\caption{(Color online) Schematic figure of current generated in response to incident electric field with tilt angle $\theta$.}
\label{app:Fig2}
\end{center}
\end{figure}

Let us consider the optical conductivity in response to an incident electric field ${\bm E}(t) = E(\omega)e^{-i\omega t}(\cos\theta,\sin\theta)$. The current generated in the direction of the electric field is expressed by ${\cal J}(t)={\cal J}_{x}(t)\cos\theta+{\cal J}_{y}(t)\sin\theta$ (see Fig.~\ref{app:Fig2}). Accordingly, the longitudinal optical conductivity $\sigma(\omega)$ is defined by ${\cal J}(\omega)=\sigma(\omega)E(\omega)$ and can be calculated from the Kubo formula as
\begin{align}
\sigma(\omega) &=\frac{1}{\hbar\omega}\sum_{\bm k} \int_{0}^{\infty} dt \left<[j^I_{\bm{k}}(t),j^I_{\bm{k}}(0)]\right>_{0}e^{i\omega t-\delta t} , \\
j^I_{\bm{k}}(t) &= j^I_{\bm{k},x}(t) \cos \theta + j^I_{\bm{k},y}(t) \sin \theta .
\end{align}
Using Eq.~(\ref{Kubo1}), we obtain
\begin{align}
\sigma(\omega)=\sigma_{xx}(\omega)\cos^{2}\theta+\sigma_{yy}(\omega)\sin^{2}\theta+\sigma_{xy}(\omega)\cos\theta\sin\theta+\sigma_{yx}(\omega)\cos\theta\sin\theta.
\label{sigmaomegaexp}
\end{align}

The current operator in the interaction picture is directly calculated as
\begin{align}
j^I_{{\bm{k}},i}(t)=\alpha_i({\bm{k}}) \left[e^{\dagger}_{\bm{k}}(t)e_{\bm{k}}(t)+h^{\dagger}_{-\bm{k}}(t)h_{-\bm{k}}(t)-1\right]+\beta_i({\bm{k}})\left[h_{-\bm{k}}(t)e_{\bm{k}}(t)+e^{\dagger}_{\bm{k}}(t)h^{\dagger}_{-\bm{k}}(t)\right],
\end{align}
with $e_{\bm{k}}(t)=e^{-i\epsilon_{\bm{k}}t} e_{\bm{k}}$ and $h_{-\bm{k}}(t)=e^{-i\epsilon_{\bm{k}}t} h_{-\bm{k}}$, where $\epsilon_{\bm k} = |f({\bm k})|$ is the band dispersion in the absence of the incident light. Substituting these equations into Eq.~(\ref{Kubo1}) and performing integration, we obtain
\begin{equation}
\sigma_{ij}(\omega)=\sum_{\bm k}\frac{\beta_i({\bm{k}})\beta_j({\bm{k}})}{\hbar\omega} \left[\frac{i(\omega-2\epsilon_{\bm{k}})+\delta}{(\omega-2\epsilon_{\bm{k}})^2+\delta^2}-\frac{i(\omega+2\epsilon_{\bm{k}})+\delta}{(\omega+2\epsilon_{\bm{k}})^2+\delta^2}\right].
\label{Absorption0}
\end{equation} 
By taking a real part of the optical conductivity, supposing $\delta \to 0$, and using Eq.~(\ref{sigmaomegaexp}), we finally obtain 
\begin{align}
{\rm{Re}}\left[\sigma(\omega)\right]
= \frac{\pi}{\hbar\omega}\sum_{\bm{k}}\left(\beta_x({\bm{k}})\cos\theta+\beta_y({\bm{k}})\sin\theta\right)^2\left[\delta(\omega-2\epsilon_{\bm{k}})-\delta(\omega+2\epsilon_{\bm{k}})\right].&
\label{Absorption}
\end{align}
The influence of the shear strain can be identified from information on $\beta_{i}({\bm k})$ that includes the dipole moment $d_i({\bm{k}})$ of graphene (see Eq.~(\ref{AB})) as well as the band structure $\epsilon_{\bm{k}}$ modified by the shear strain. The real part of the optical conductivity can be regarded as the absorption coefficient as long as the condition $\rm{Im}[\sigma(\omega)] \ll \rm{Re}[\sigma(\omega)]$ is satisfied. As is written in the standard textbook~\cite{haug2009quantum}, this condition holds for typical semiconductors. By using this formula, we can calculate its polarization angle dependence as a function of the shear strain parameter $\zeta$. In our numerical calculation, we performed the integration of $\bm{k}$ in Eq.(\ref{Absorption0}) with $\delta=0.01/\omega_{0}$, where a finite value of $\delta$ indicates homogeneous broadening of the system.

\section{Polarization Angle Dependence of Absorption Coefficients in Graphene under Shear Strain}
\label{app:subsec4}
\begin{figure}[tb]
\begin{center}
\includegraphics[width=16cm]{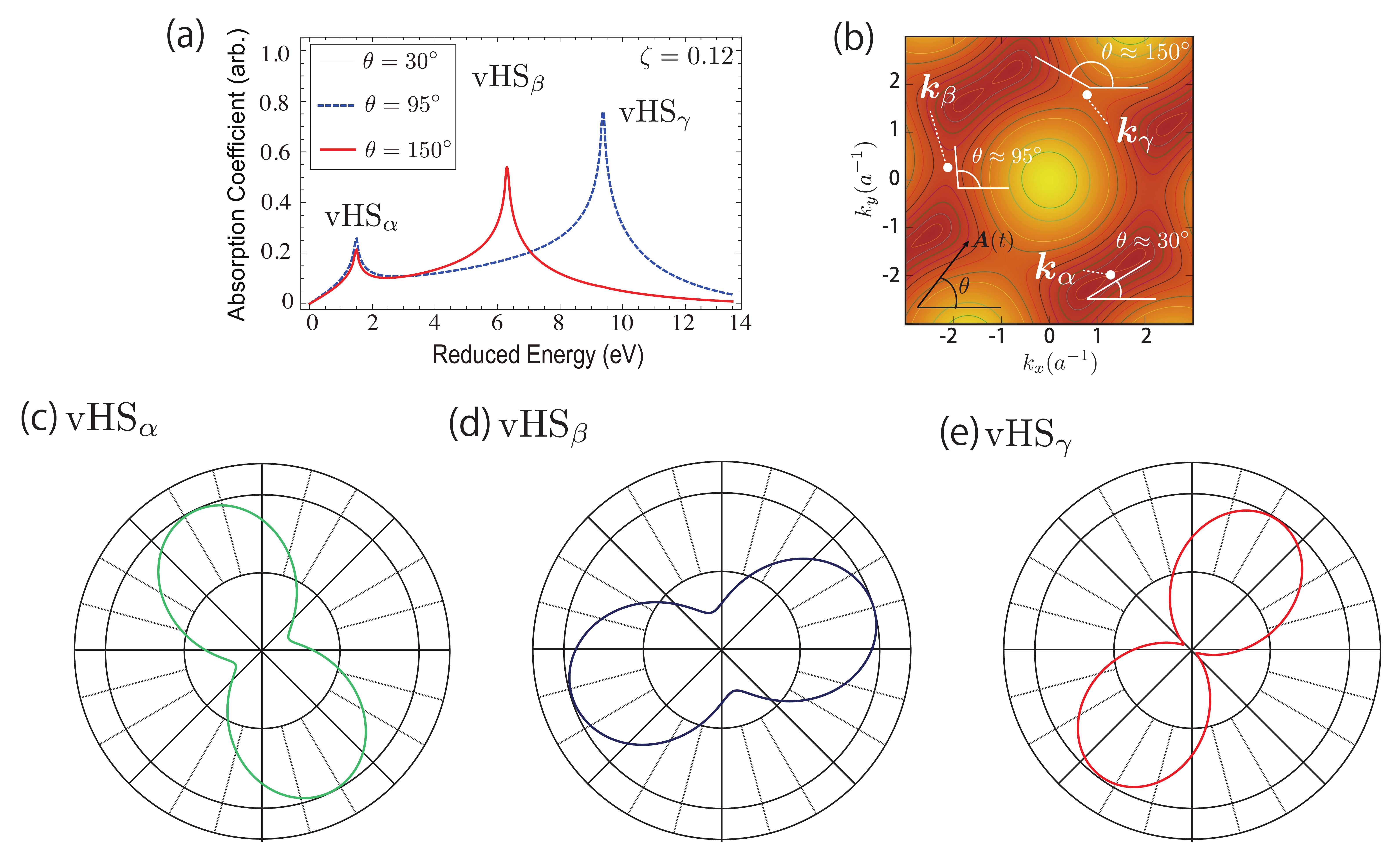}
\caption{(Color online) (a): Frequency dependence of the absorption coefficient of graphene for $\zeta=0.12$. The green, blue, and red lines are for three different angles of the incident THz light, $\theta=30^{\circ}$, $95^{\circ}$, and $120^{\circ}$. (b): Contour plot of the energy band of graphene for $\zeta=0.12$. Here, ${\bm k}_{\alpha}$, ${\bm k}_{\beta}$, and {${\bm k}_{\gamma}$} indicate the positions of the saddle points, which induce ${\rm vHS}_\alpha$, ${\rm vHS}_\beta$, and ${\rm vHS}_\gamma$, respectively. The azimuth angle of the band minima (the Dirac cones) measured from the saddle points are given as $\theta \approx 30^{\circ}$, $95^{\circ}$, and $120^{\circ}$. (c)-(e): Polarization angle dependence of the absorption coefficients at ${\rm vHS}_{\alpha}$, ${\rm vHS}_{\beta}$, and ${\rm vHS}_\gamma$, respectively. }
\label{app:Fig3}
\end{center}
\end{figure}

In this section, we provide additional information on the polarization angle dependence of the absorption coefficients. Fig.~\ref{app:Fig3}~(a) show the frequency dependence of the absorption coefficient for different incident light angles $\theta=30^{\circ}$ (green line), $\theta=95^{\circ}$ (blue line), and $\theta=120^{\circ}$ (red line) in the case of $\zeta=0.12$. The absorption coefficient has three peaks, which correspond to three van Hove singularities (vHS) resolved by the shear strain. In addition, the absorption coefficient has different polarization angle dependences near these three peaks. In the main text, these three vHSs are labeled ${\rm vHS}_{\alpha}$, ${\rm vHS}_{\beta}$, and ${\rm vHS}_{\gamma}$, respectively. Fig.~\ref{app:Fig3}~(b) shows the contour plot of the energy dispersion $\epsilon_{\bm k}$ of the conduction band for $\zeta=0.12$. Here, the three vHSs are induced from the three saddle points of the band dispersion, whose wavenumbers are denoted by ${\bm k}_i$ ($i=\alpha,\beta,\gamma$). The polarization angle dependences of the absorption coefficients at the three peaks are plotted in Fig.~\ref{app:Fig3}~(c)-(e). We can see that the absorption coefficients reach maximum values at $\theta\approx 120^{\circ}$, $\theta \approx 5^{\circ}$, and $\theta \approx 60^{\circ}$ for ${\rm vHS}_{\alpha}$, ${\rm vHS}_{\beta}$, and ${\rm vHS}_{\gamma}$, respectively. These orientations are approximately perpendicular to the directions of the band minima (the Dirac points) measured from the saddle points indicated by the white lines in Fig.~\ref{app:Fig3}~(b). These results clearly show that the polarization angle varies depending on the energy range of the incident light, and the maximum and minimum angles are determined by the corresponding vHSs.

\bibliography{SuppleGrapheneHHGref.bib}